\documentclass{appolb}
\usepackage{graphicx}
\usepackage{amsmath}
\usepackage{amsfonts}
\usepackage{bm}

\begin{document}
	\title{Free $\mu^{+}\mu^{-}$ and bound-free $e^{+}e^{-}$ pair production in relativistic heavy-ion collisions
	}%

\author{Melek YILMAZ~\c{S}ENG\"{u}L
	\address{Department of Physics, Faculty of Science and Letters, {I}stanbul K\"{u}lt\"{u}r University, {I}stanbul,TURKEY}
	\address{melek.sengul@iku.edu.tr, melekaurora@yahoo.com}
}
\maketitle
\begin{abstract}
	
In our previous studies, we calculated the cross-section for the production of free $\mu^{+}\mu^{-}$ pairs using a probability function dependent on the impact parameter, and also, we computed bound-free $e^{+}e^{-}$ pair production cross section. In this study, unlike the others, we obtained the cross section for the production of bound-free $e^{+}e^{-}$ pairs using the probability function dependent on the impact parameter. Here also, we simultaneously calculated the cross sections for producing free $\mu^{+}\mu^{-}$ with bound-free $e^{+}e^{-}$ pairs in relativistic $Pb+Pb$ collisions at the LHC with Lorentz factors at $\gamma=1500$ and $\gamma=3000$ reported in the literature. Our calculation is based on second-order perturbation theory, implemented by using Monte Carlo integration techniques. We present impact parameter dependent pair production probabilities for free $\mu^{+}\mu^{-}$ and bound-free $e^{+}e^{-}$ pair production processes. Notably, the bound-free $e^{+}e^{-}$ pair production was evaluated by using a formalism adapted from that of free $\mu^{+}\mu^{-}$ pair production. Our results obtained here for the free $\mu^{+}\mu^{-}$ with bound-free $e^{+}e^{-}$ pair production cross sections are compared with those from other theoretical methods available in the literature. This study aims to provide predictions that can be tested against future experimental data, thereby offering a valuable check for quantum electrodynamics (QED) in strong electromagnetic fields.

\end{abstract}

\section{Introduction}
In ultra-peripheral heavy ion collisions at the LHC, free lepton pairs are produced via the intense electromagnetic fields of the colliding nuclei. The corresponding cross sections scale as $(Z\alpha)^4 ln^3(\gamma)$, where $\alpha$ is the fine structure constant and $\gamma$ is the Lorentz factor\cite{1,2,3}. In a subset of these processes, the produced negatively charged lepton can be captured by one of the ions. This capture process leads to beam depletion, alters the ion charge state, and ultimately causes the affected ion to be lost from the beam pipe. The cross section for this bound-free pair production (BFPP) scales as $(Z\alpha)^8 ln(\gamma)$ \cite{3,4,5,6}. Comparing the scaling behaviors, the additional factor of an extra third power of heavy ion for BFPP relative to free pair production originates from the bound-state wave function of the captured electron, while the remaining  power of heavy ion factor arises from the normalization constant of the positron wave function.

In order to represent a final state pair production cross section, 
the cross section integral can be expressed over the "photon fluxes" of each of the ions times the on-shell two photon to the final cross section \cite{glu}.

In this study, we investigate two types of particle creation that occur when ultra-fast lead ions collide. We focus on the production of free $\mu^{+}\mu^{-}$ and bound-free $e^{+}e^{-}$ pairs. We calculated the probability (cross-section) for these processes to occur in $Pb+Pb$ collisions at the LHC. Specifically, we analyzed collisions at two different, ultra-relativistic velocities, corresponding to Lorentz factors of $\gamma=1500$ and $\gamma=3000$.

This work investigates a combined theoretical approach to two quantum electrodynamic (QED) processes occurring in the strong electromagnetic fields of ultra relativistic heavy-ion collisions: bound-free  $e^{+}e^{-}$ pair production and free $\mu^{+}\mu^{-}$ pair production. We present the cross section of the production of free $\mu^{+}\mu^{-}$ and bound-free $e^{+}e^{-}$ pairs in relativistic $Pb+Pb$ collisions at the LHC moving with Lorentz factors $\gamma=1500$ and $\gamma=3000$. In the context of the LHC, these specific final states can be generated via two fundamental interaction mechanisms. The electrons may be captured by the same ions as in Eqs.~(\ref{1}) and ~(\ref{2})
\begin{eqnarray}
	\label{1}
	Z_x + Z_y &\rightarrow& \: (Z_x+e^-)_{1s}\:+\: Z_y \:+\:\mu^{+}\:+\:\mu^{-}\:+\: e^+
\end{eqnarray}
or
\begin{eqnarray}
	\label{2}
	Z_x+Z_y &\rightarrow& \: Z_x \:+\:\mu^{+}\:+\:\mu^{-}\:+\: e^+\:+\:(Z_y+e^-)_{1s}.
\end{eqnarray}

By doing these calculations, we are trying to take a new point of view to the strong field in QED.

In our analysis, we employ natural units (where $\hbar=c=m=e=1$), a standard convention for ultra relativistic systems. The separation between the centers of the colliding nuclei is defined as the impact parameter, denoted by $b$.

\section{Formalism}
In this study, we present a detailed analysis of multiple pair production, including both capture and free lepton pair production. Our calculations are performed within the framework of quantum electrodynamics (QED) by using lowest-order perturbation theory \cite{mys}.
To calculate the cross sections for the production of free $\mu^{+}\mu^{-}$ pairs and bound-free $e^{+}e^{-}$ pairs, we first characterized the pair production dynamics for each process individually. Subsequently, using Eq.~(\ref{3}) provided below, we computed the simultaneous production of free $\mu^{+}\mu^{-}$ pairs alongside bound-free $e^{+}e^{-}$ pairs.

\begin{eqnarray}
	\label{3}
	\sigma_{bfpp+\mu^{+}\mu^{-}}=\int^{\infty}_{2R}P_{bfpp}(b)P_{\mu^{+}\mu^{-}}(b)2\pi\:b\:db,
\end{eqnarray}

where $2R$ represents is the minimum impact parameter for the ultra-peripheral heavy ion collisions and $R$ is the radius of the $Pb$ ion that is approximately equal to $6.62 fm$ \cite{mys1,mysg}.
First, we present the detailed calculations for the probability of free $\mu^{+}\mu^{-}$ pair production in relativistic $Pb+Pb$ collisions at the LHC, with Lorentz factors of $\gamma=1500$ and $\gamma=3000$. We employ a methodology similar to that used in our previous work \cite{mys1} to derive the probability expression for these different energy regimes.

The cross section for free $\mu^{+}\mu^{-}$ pair production, which depends on the impact parameter and includes contributions from both direct and crossed terms in second-order perturbation theory, can be represented as:

\begin{equation} \label{4}
	\sigma_{\mu^{+}\mu^{-}}=\int d^2 b \sum_{k_1>0} \, \sum_{q_1<0} \, \big|\langle \chi_{k_1}^{(+)} | S | \chi_{q_1}^{(-)} \rangle \big|^2.
\end{equation}

Here, the summation over $k_1$ and $q_1$ corresponds to states outside and inside the Dirac sea, respectively. Anti-particle states are denoted by $|\chi_{k_1}^{(+)} \rangle$, and particle states by $|\chi_{q_1}^{(-)}\rangle$. The scattering matrix $S=S_{ab}+S_{ba}$, which includes both direct and crossed terms, is expressed as a series expansion.

Muons, being heavy leptons, have a Compton wavelength that is smaller than the radii of heavy ions ($Au$, $Pb$). Therefore, in such collisions, the Wood–Saxon form factor of nucleons cannot be neglected \cite{mel1,mel2}. In our calculations, we employed the two-parameter Fermi (2pF) function—also known as the Wood–Saxon distribution—to model the nuclear charge density \cite{baltz2}. This functional form is given by

\begin{equation} \label{5}
	\rho(r)= \frac{\rho_0}{1+\exp \big( \frac{r-R}{a}\big)}
\end{equation}   

where $R$ is the nuclear radius, $a$ is the surface diffuseness (related to the thickness of the nuclear surface), and $\rho_0$ is determined through normalization.

As detailed in our previous work \cite{mys1}, the total cross section for $\mu^{+}\mu^{-}$ pair production is given by

	\begin{eqnarray}\label{6}
    \sigma_{\mu^{+}\mu^{-}} & = &
	\frac{1}{4 \beta^2} \sum_{{\sigma_{k1}}} \sum_{\sigma_{p1}} \int \frac{d^2 k_{1\perp} d k_{1z} d^2 q_{1 \perp} d q_{1z} d^2 p_{1\perp}}{(2 \pi)^8} \times 
	\nonumber \\
	& & |\mathbb{A}_1^{(+)}(k_1,q_1; \mathbf{p}_{1\perp})+ \mathbb{A}_1^{(-)}(k_1,q_1; \mathbf{k}_{1\perp}+\mathbf{q}_{1 \perp}-\mathbf{p}_{1\perp})|^2
\end{eqnarray}

where,

\begin{subequations}
\begin{eqnarray}
		\mathbb{A}_1^{(+)}(k_1,q_1; \mathbf{p}_{1\perp}) & = &
		F_1(\mathbf{k}_{1\perp}-\mathbf{p}_{1\perp}; \omega_{1\mathcal{A}}) \times 	\nonumber \\
		& & 
		F_1(\mathbf{p}_{1\perp}-\mathbf{q}_{1 \perp}; \omega_{1\mathcal{B}}) \mathcal{T}_{k_1q_1}(\mathbf{p}_{1\perp}; +\beta)
\end{eqnarray}

and

\begin{eqnarray}
		\mathbb{A}_1^{(-)}(k_1,q_1; \mathbf{p}_{1\perp}) & = &
		F_1(\mathbf{k}_{1\perp}-\mathbf{p}_{1\perp}; \omega_{1\mathcal{B}})
		\times 	\nonumber \\
		& &
		F_1(\mathbf{p}_{1\perp}-\mathbf{q}_{1\perp}; \omega_{1\mathcal{A}}) \mathcal{T}_{k_1q_1}(\mathbf{p}_{1\perp};-\beta).
\end{eqnarray}
\end{subequations}

In the equations above, the functions $\mathcal{T}_{k_1q_1}$ represent the reduced Feynman amplitudes, while $F_1(\mathbf{k}_{1\perp}-\mathbf{p}_{1 \perp};\omega_{1\mathcal{A},1\mathcal{B}})$ and  $F_1(\mathbf{p}_{1\perp}-\mathbf{q}_{1 \perp};\omega_{1\mathcal{A},1\mathcal{B}})$ correspond to the scalar components of the electromagnetic fields generated by the heavy ions, expressed in momentum space. Here, $\mathbf{k}_{1\perp}$ and $\mathbf{q}_{1\perp}$ denote the transverse momenta, and $k_{1z}$ and $q_{1z}$ the longitudinal momenta of the produced lepton pairs, whereas  $\mathbf{p}_{1\perp}$ is the transverse momentum of the intermediate states. The frequencies of the colliding heavy ions are denoted by $\omega_{1\mathcal{A}}$ and $\omega_{1\mathcal{B}}$. 

The explicit form of these scalar fields in momentum space can be expressed as

\begin{subequations}
\begin{eqnarray}\label{12}
	F_1(\mathbf{k}_{1\perp}-\mathbf{p}_{1\perp}; \omega_{1\mathcal{A}})=
	\frac{4\pi Z\gamma^2\beta^2}{
		\left(\omega^{2}_{1\mathcal{A}} + \gamma^2\beta^2(\mathbf{k}_{1\bot} -
		\mathbf{p}_{1\bot})^2\right)} \,
\end{eqnarray}
and
\begin{eqnarray}\label{13}
	F_1(\mathbf{p}_{1\bot}-\mathbf{q}_{1\bot};\omega_{1\mathcal{B}}) & = &
	\frac{4\pi Z\gamma^2\beta^2}{
		\left(\omega^{2}_{1\mathcal{B}} + \gamma^2\beta^2(\mathbf{p}_{1\bot} -
		\mathbf{q}_{1\bot})^2\right)}. \,
\end{eqnarray}
\end{subequations}
The reduced Feynman amplitude can be represented as

\begin{eqnarray}\label{14}
	\mathcal{T}_{k_1q_1}(\mathbf{p}_{1\bot};+\beta) 
	& =& \sum_s \sum_{\sigma_{p_1}}
	\frac{1}{\left(E^{(s)}_{p_1} - \left(\frac{E^{(+)}_{k_1}+E^{(-)}_{q_1}}{2}\right) 
		+\beta(\frac{k_{1z}-q_{1z}}{2})\right)} 
	\nonumber \\[0.2cm]
	&   & 
	\left\langle\textbf{u}^{(+)}_{\sigma_{k_1}}\left|{(1-\beta\alpha_z)}\right|{\textbf{u}^{(s)}_{\sigma_{p_1}}}\right\rangle
	\left\langle\textbf{u}^{(s)}_{\sigma_{p_1}}\left|{(1+\beta\alpha_z)}\right|{
		\textbf{u}^{(-)}_{\sigma_{q_1}}}\right\rangle.
\end{eqnarray}
It depends explicitly on the velocity of the heavy ions ($\beta$) , ($\mathbf{p}_{1\bot}$) ,($p_{1z}$) longitudinal momentum of the intermediate states and ($q_1$) anti-muon momentum. $\textbf{u}^{(s)}_{\sigma_{p1}}$ represents the spinor part of the intermediate states \cite{glu,gu}.

When describing the impact parameter dependent cross section for $\mu^{+}\mu^{-}$ pair production in relativistic heavy ion collisions, the Bessel functions involved are highly oscillatory, particularly for large impact parameters. This behavior presents significant challenges for the numerical evaluation of the cross section. To address this, the expression in Eq. \eqref{8} can be separated into two parts.

\begin{eqnarray}\label{8}
	\frac{d\sigma_{\mu^{+}\mu^{-}}}{d b}= \int_0^{\infty} q_1 dq_1 \, b J_0(q_1b) \, \mathcal{F}_1(q_1)
\end{eqnarray}

Here, $\mathcal{F}_1(q_1)$ is a nine-dimensional integral. For a fixed value of $q_1$, it was evaluated using the Monte Carlo integration method \cite{glu,gu}.

\begin{eqnarray}\label{15}
	\mathcal{F}_1(q_1)
	&=&\frac{\pi}{8 \beta^2} \sum_{\sigma_{k_1}} \sum_{\sigma_{q_1}} \int_0^{2 \pi} d \phi_{q_1} \int \frac{dk_{1z} dq_{1z}  d^2 k_{1\perp} d^2 K_1 d^2 Q_1}{(2 \pi)^{10}}\times \nonumber \\ 
	&&
	\bigg\{ F_1 \Big( \frac{\mathbf{Q}_1-  \mathbf{q_1}}{2}; \omega_{1\mathcal{A}} \Big) F_1 \big(-\mathbf{K}_1;\omega_{1\mathcal{B}} \big) \mathcal{T}_{k_1q_1} \Big(\mathbf{k}_{1\perp}-\frac{\mathbf{Q}_1-\mathbf{q_1}}{2}; \beta \Big) \nonumber \\
	&&
	+F_1 \Big( \frac{\mathbf{Q}_1-  \mathbf{q_1}}{2}; \omega_{1\mathcal{A}} \Big) F_1 \big(-\mathbf{K}_1;\omega_{1\mathcal{B}} \big) \mathcal{T}_{k_1q_1} \big(\mathbf{k}_{1\perp}- \mathbf{K}_1; -\beta \big) \bigg\}\times \nonumber \\
	&&
	\bigg\{F_1 \bigg( \frac{\mathbf{Q}_1 + \mathbf{q_1}}{2}; \omega_{1\mathcal{A}} \bigg) F_1 \big(-\mathbf{q_1}- \mathbf{K_1};\omega_{1\mathcal{B}} \big) \mathcal{T}_{k_1q_1} \bigg(\mathbf{k}_{1\perp}-\frac{\mathbf{Q}_1+ \mathbf{q_1}}{2}; \beta \bigg) \nonumber \\ 
	&&
	+F_1 \Big(\frac{\mathbf{Q}_1+  \mathbf{q_1}}{2}; \omega_{1\mathcal{A}} \Big) F \big(-\mathbf{q_1}- \mathbf{K}_1;\omega_{1\mathcal{B}} \big) \mathcal{T}_{k_1q_1} \big(\mathbf{k}_{1\perp}+\mathbf{q_1}- \mathbf{K}_1; -\beta \big) \bigg\}.
	\end{eqnarray}

For these calculations, $\mathbf{K_1}$ and $\mathbf{Q_1}$ are new variables, defined as functions of $\mathbf{q_1}$ and $\mathbf{p_1}$ \cite{glu,gu}. 

These new variables can be expressed as;

\begin{subequations}
\begin{equation} 
	\mathbf{Q}_{1}= 2\mathbf{k}_{1\perp}-\mathbf{p}_{1\perp}-\mathbf{p}_{1\perp}^{'} 
\end{equation}

\begin{equation} 
		\mathbf{q}_{1}= \mathbf{p}_{1\perp}-\mathbf{p}_{1\perp}^{'}
\end{equation}

\begin{equation} 
	\mathbf{K}_{1}= -\mathbf{p}_{1\perp}+\mathbf{q}_{1\perp}
\end{equation}
\end{subequations}

Consequently, $\mathcal{F}_1(q_1)$ can be expressed explicitly as shown in the equation above. The cross section for free $\mu^{+}\mu^{-}$ pair production can then be formulated as a function of $\mathcal{F}_1(q_1)$, which encapsulates the scalar components of the electromagnetic fields and depends on the fixed value of the anti-muon momentum $q_1$,

\begin{equation} \label{171}
	\mathcal{F}_1(q_1)= \mathcal{F}_1(0)  e^{-a_1\sqrt{q_1}}=\sigma_{\mu^{+}\mu^{-}}e^{-a_1\sqrt{q_1}}.
\end{equation}

The function $\mathcal{F}_1(q_1)$ was computed by numerically integrating Eq.\eqref{171} using Monte Carlo techniques for several fixed values of  $q_1$. The results of these calculations, corresponding to different  $q_1$ values, are shown for $\gamma=1500$ in Fig.~\ref{1} and for $\gamma=3000$ in Fig.~\ref{2}. As shown in Fig.~\ref{1} and Fig.~\ref{2}, the resulting dependence of $\mathcal{F}_1(q_1)$ on $q_1$ follows an exponential trend, which is best described by the functional form $e^{-a_1\sqrt{q_1}}$. Utilizing this exponential relationship for $\mathcal{F}_1(q_1)$, the probability $P_{\mu^{+}\mu^{-}}(b)$ for producing a single $\mu^{+}\mu^{-}$ pair from second-order Feynman diagrams can be expressed as follows,

\begin{equation} \label{18}
	P_{\mu^{+}\mu^{-}}(b)=\frac{1}{2\pi b} \frac{d\sigma}{d b}=\frac{1}{2\pi}\mathcal{F}_1(0)\int_0^{\infty} q_1dq_1 \, J_0(q_1b) \, e^{-a_1\sqrt{q_1}}.
\end{equation}

The parameter $a_1$ is a constant, found to be $7.223\lambda_{C}^{\mu}$ at $\gamma=1500$ and $8.4\lambda_{C}^{\mu}$ at $\gamma=3000$. It is proportional to the Compton wavelength of the muon ($\lambda_{C}^{\mu}$) and independent of both the collision energy and the nuclear charge. Similar calculations for other energies are reported in \cite{mys1}. Consequently, the function $\mathcal{F}_1(q_1)$ can be expressed in the form $\mathcal{F}_1(q_1)=\mathcal{F}_1(0)e^{-a_1\sqrt{q_1}}$ where $\mathcal{F}_1(0)$ represents the total cross section at $q_1=0$. This value, calculated via the Monte Carlo integration method, corresponds to $1.78$ $mbarn$ for $\sigma_{\mu^{+}\mu^{-}}$ at $\gamma=1500$ and $2.564$ $mbarn$ at $\gamma=3000$.

\begin{figure}
	\includegraphics[scale=0.7]{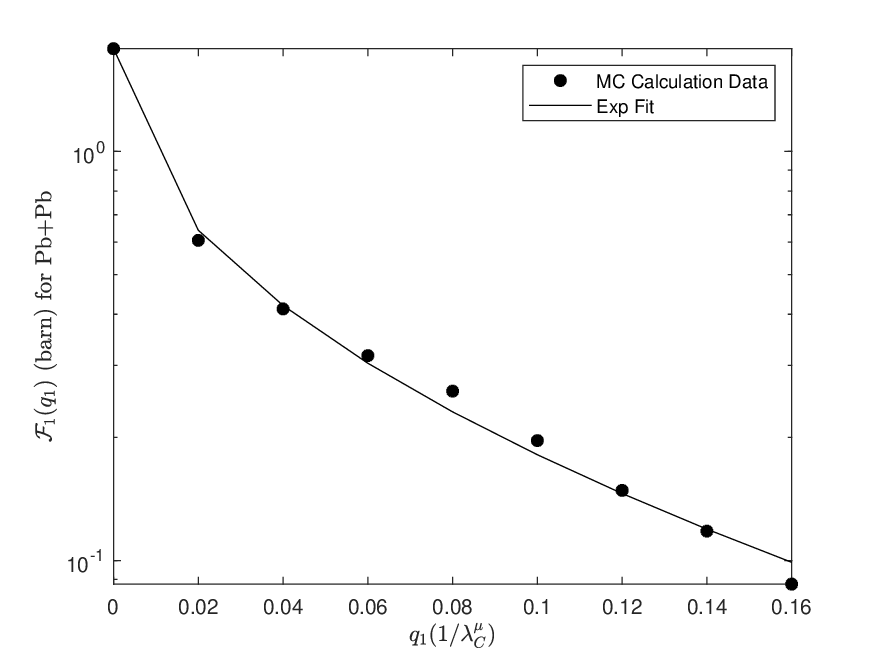}
	\caption{The function $\mathcal{F}_1(q_1)$ for $\mu^{+}\mu^{-}$ pair production is calculated as a function of $q_1$ for $Pb+Pb$ collisions at $\gamma=1500$ in the center of momentum frame. The points give the Monte Carlo calculations results for different  $q_1$ values and the smooth curve represents our fit function for these points.}
	\label{1}
\end{figure}

\begin{figure}
	\includegraphics[scale=0.7]{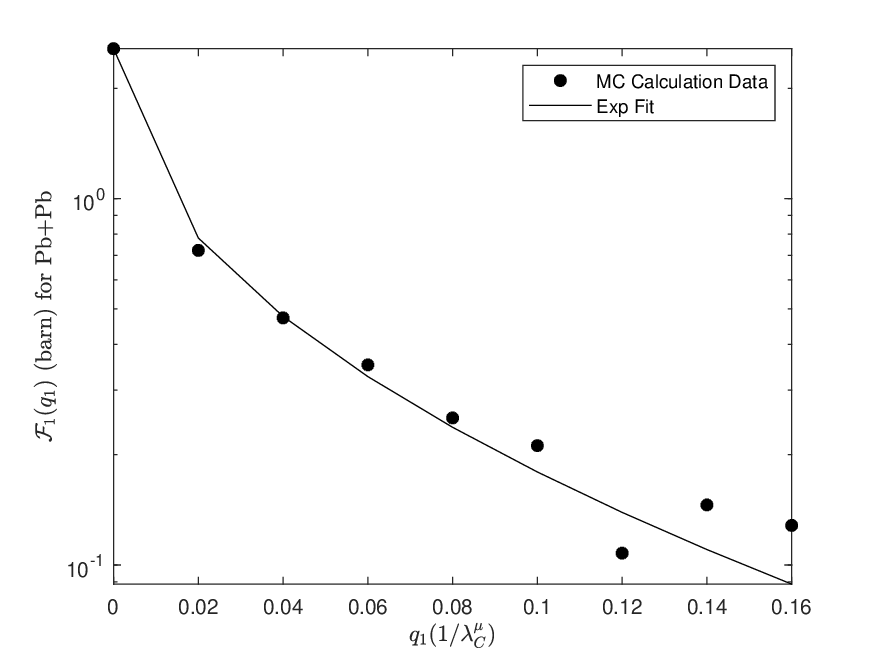}
	\caption{The function $\mathcal{F}_1(q_1)$ for $\mu^{+}\mu^{-}$ pair production is calculated as a function of $q_1$ for $Pb+Pb$ collisions at $\gamma=3000$ in the center of momentum frame. The points give the Monte Carlo calculations results for different  $q_1$ values and the smooth curve represents our fit function for these points.}
	\label{2}
\end{figure}

The $\mu^{+}\mu^{-}$ pair production probability as a function of impact parameter $b$ is shown in Figs.\ref{3} and \ref{4} for $\gamma=1500$ and $\gamma=3000$, respectively. Its behavior closely matches the impact parameter dependent probability profiles calculated using the Equivalent Photon Method in \cite{15} for the same Lorentz factors. 

\begin{figure}
	\includegraphics[scale=0.7]{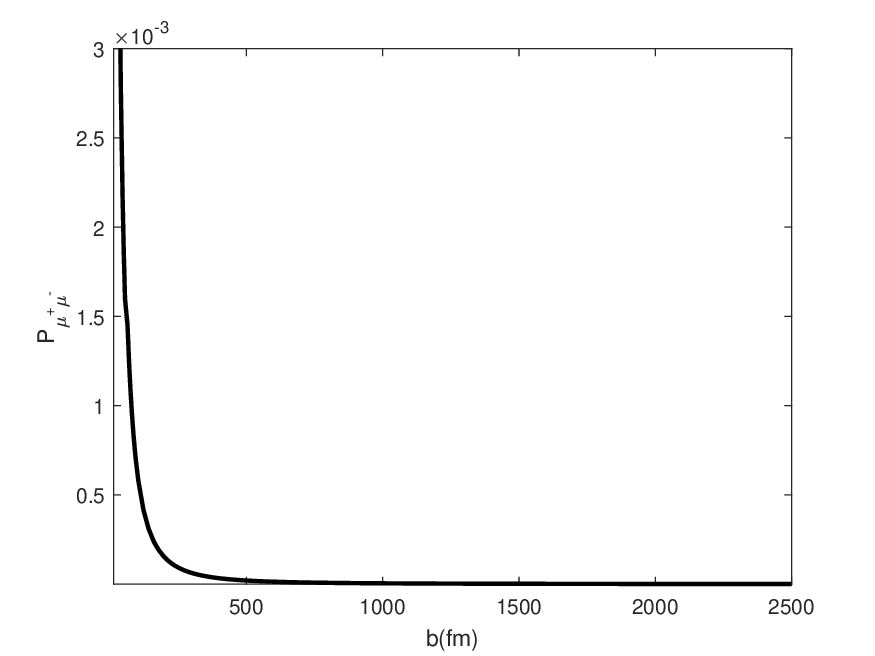}
	\caption{The probability of $\mu^{+}\mu^{-}$ pair production is shown as a function of the impact parameter for $Pb+Pb$ collisions at $\gamma=1500$ in the center of momentum frame.}
	\label{3}
\end{figure}

\begin{figure}
	\includegraphics[scale=0.7]{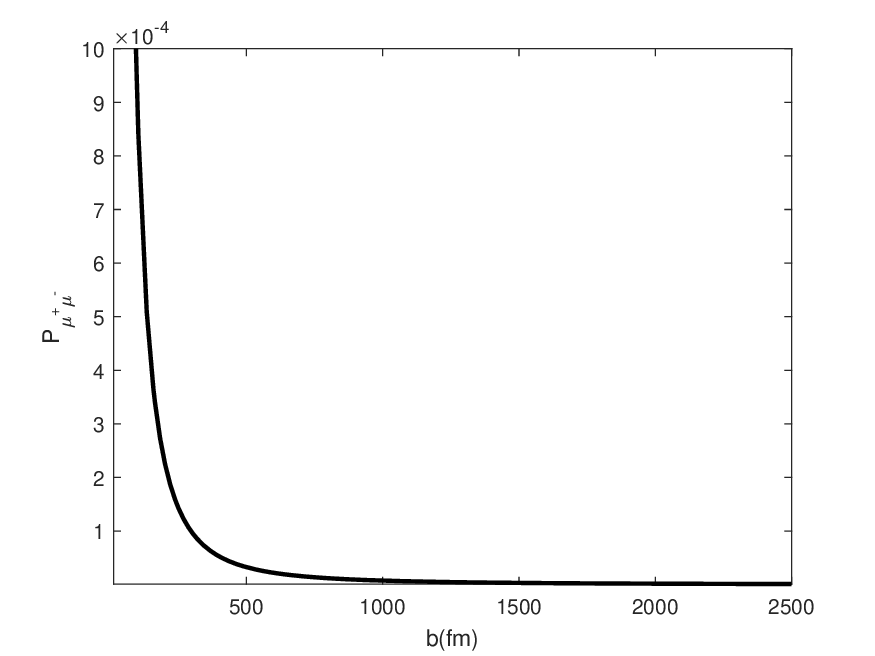}
	\caption{The probability of $\mu^{+}\mu^{-}$ pair production is shown as a function of the impact parameter for $Pb+Pb$ collisions at $\gamma=3000$ in the center of momentum frame.}
	\label{4}
\end{figure}

In the second step, the cross section for bound-free $e^{+}e^{-}$ pair production is derived, where the direct and crossed terms are characterized by the corresponding Feynman diagrams at the lowest order of QED,

\begin{eqnarray}\label{9}
	\sigma_{bfpp} & = & 
	\int d^2b\sum_{q<0}\left|\left\langle 
	\Psi^{(-)}\left|S\right|\Psi^{(+)}_q\right\rangle\right|^2.
\end{eqnarray}

To compute these terms, the free positron wave function, $\Psi^{(+)}_q$, is described by the Sommerfeld-Maue wave function, while the captured electron wave function, $\Psi^{(-)}$, is given by the Darwin wave function. Within second-order perturbation theory, the bound-free pair production (BFPP) cross section can be expressed as \cite{3}

\begin{eqnarray}\label{9a}
	\sigma_{bfpp} & = & 		   
	\frac{\left|N_{+}\right|^{2}}{4\beta^2} \,
	\frac{1}{\pi} \left(\frac{Z}{a_H}\right)^3
	\sum_{\sigma_q} \int \frac{d^3qd^2p_\bot}{(2\pi)^5}\times 
	\nonumber \\
	& & | \mathbb{A}^{(+)}(q; \mathbf{p}_\bot) + 
	\mathbb{A}^{(-)}(q; \mathbf{q}_\bot  - \mathbf{p}_\bot)|^2 ,
\end{eqnarray}
with

\begin{subequations}
\begin{eqnarray}
	\mathbb{A}^{(+)}(q ;\mathbf{p}_\bot) 
	&=&
	F(-\mathbf{p}_\bot; \omega_{\mathcal{A}}) \, 
	F(\mathbf{p}_\bot-\mathbf{q}_\bot; \omega_{\mathcal{B}}) \,
	\mathcal{T}_q(\mathbf{p}_\bot;+\beta),				
\end{eqnarray}

and

\begin{eqnarray}\label{11}
	\mathbb{A}^{(-)}(q ;\mathbf{q}_\bot-\mathbf{p}_\bot) 
	&=&
	F(\mathbf{p}_\bot-\mathbf{q}_\bot; \omega_{\mathcal{B}}) \,
	F(-\mathbf{p}_\bot; \omega_{\mathcal{A}}) \mathcal{T}_q(\mathbf{q}_\bot-\mathbf{p}_\bot; -\beta). \, 
\end{eqnarray}
\end{subequations}

In the equations above, $\mathcal{T}_q$ couples the intermediate photon lines to the outgoing electron and positron lines. It depends explicitly on the velocity of the heavy ions ($\beta$), ($\mathbf{p}_\bot$) transverse momentum of the intermediate states, ($p_z$) longitudinal momentum of the intermediate states, and ($q$) positron momentum.   $\textbf{u}^{(s)}_{\sigma_{p}}$ represents the spinor part of the intermediate-state \cite{3}. The terms $F(\mathbf{p}_\bot-\mathbf{q}_\bot; \omega_{\mathcal{B}})$ and $F(-\mathbf{p}_\bot; \omega_{\mathcal{A}})$ represent the scalar-field components of the two ions in momentum space. They depend on the frequencies  $\omega_{\mathcal{A}}$ and $\omega_{\mathcal{B}}$, which correspond to the colliding heavy ions \cite{3}.

The explicit form of the scalar fields in momentum space can be expressed in terms of the corresponding frequencies as

\begin{subequations}
\begin{eqnarray}
	F(-\mathbf{p}_{\bot};\omega_{\mathcal{A}})=
	\frac{4\pi Z}{\left(\frac{Z^2}{a^{2}_{H}}\,+\,
		\frac{\omega^{2}_{\mathcal{A}}}{\gamma^2\beta^2}\,+\,
		\mathbf{p}^{2}_{\bot}\right)} 	
\end{eqnarray}
and
\begin{eqnarray}
	F(\mathbf{p}_{\bot}-\mathbf{q}_{\bot};\omega_{\mathcal{B}}) & = &
	\frac{4\pi Z\gamma^2\beta^2}{
		\left(\omega^{2}_{\mathcal{B}} + \gamma^2\beta^2(\mathbf{p}_{\bot} -
		\mathbf{q}_{\bot})^2\right)}. \,
\end{eqnarray}
\end{subequations}
The intermediate photon lines can be represented as

\begin{eqnarray}\label{14}
	\mathcal{T}_q(\mathbf{p}_{\bot}:+\beta) 
	& =& \sum_s \sum_{\sigma_p}
	\frac{1}{\left(E^{(s)}_{p} - \left(\frac{E^{(-)}+E^{(+)}_{q}}{2}\right) 
		-\beta\frac{q_{z}}{2}\right)} 
	\left[1 + \frac{\bm{\alpha}\cdot\mathbf{p}}{2m}\right]
	\nonumber \\[0.2cm]
	&   & 
	\left\langle\textbf{u}\left|{(1-\beta\alpha_z)}\right|{\textbf{u}^{(s)}_{\sigma_p}}\right\rangle
	\left\langle\textbf{u}^{(s)}_{\sigma_p}\left|{(1+\beta\alpha_z)}\right|{
		\textbf{u}^{(+)}_{\sigma_q}}\right\rangle.
\end{eqnarray}

The bound-free pair production (BFPP) cross section was derived as a function of impact parameter, as shown below.

\begin{eqnarray}\label{15}
	\frac{d\sigma_{bfpp}}{db}=\int^{\infty}_{0}q dq  b J_{0}(qb){\cal F}(q).
	\label{e2}
\end{eqnarray}

This equation contains a highly oscillatory Bessel function of order zero. The function ${\cal F}(q)$ is a six-dimensional integral, defined below for a fixed value of $q$

\begin{eqnarray}\label{16}
	{\cal F}(q) &=&\frac{\pi}{8\beta^{2}}\left|N_{+}\right|^{2}\frac{1}{\pi}\left(\frac{Z}{a_{H}}\right)^{3}
	\sum_{\sigma_{q}}\int^{2\pi}_{0}d\phi_{q}\int\frac{dq_{z}d^{2}Kd^{2}Q}{(2\pi)^{7}} \times
	\nonumber \\
	&&\lbrace F\left[\frac{1}{2}(\mathbf{Q}-\mathbf{q});\omega_{\mathcal{A}}\right]F\left[\mathbf{-K};\omega_{\mathcal{B}}\right]\mathcal{T}_{q}\left[-\frac{1}{2}(\mathbf{Q}-\mathbf{q});\beta\right]\nonumber \\
	&&
	+F\left[\mathbf{-K};\omega_{\mathcal{B}}\right]F\left[\frac{1}{2}(\mathbf{Q}-\mathbf{q});\omega_{\mathcal{A}}\right]\mathcal{T}_{q}\left[\mathbf{K};-\beta\right]\rbrace \times \nonumber \\
	&& \lbrace F\left[\frac{1}{2}(\mathbf{Q}+\mathbf{q});\omega_{\mathcal{A}}\right]F\left[\mathbf{-K};\omega_{\mathcal{B}}\right]\mathcal{T}_{q}\left[-\frac{1}{2}(\mathbf{Q}+\mathbf{q});\beta\right]\nonumber \\
	&&
	+F\left[\mathbf{-K};\omega_{\mathcal{B}}\right]F\left[\frac{1}{2}(\mathbf{Q}+\mathbf{q});\omega_{\mathcal{A}}\right]\mathcal{T}_{q}\left[\mathbf{K};-\beta\right]\rbrace.
	\label{fq}
\end{eqnarray}

For these calculations, $\mathbf{K}$ and $\mathbf{Q}$ are defined as new variables, each a function of $\mathbf{q}$ and $\mathbf{p}$ \cite{3}.

These new variables can be expressed as;

\begin{subequations}
\begin{equation} 
	\mathbf{Q}= -\mathbf{p}_{\perp}-\mathbf{p}_{\perp}^{'} 
\end{equation}

\begin{equation} \label{161b}
	\mathbf{q}= \mathbf{p}_{\perp}-\mathbf{p}_{\perp}^{'}
\end{equation}

\begin{equation} \label{161c}
	\mathbf{K}= -\mathbf{p}_{\perp}+\mathbf{q}_{\perp}
\end{equation}
\end{subequations}

Following the numerical integration of Eq.~(\ref{16}), a simple analytical form for ${\cal F}(q)$ is obtained for a fixed $q$, as expressed in the equation below,

\begin{eqnarray}\label{17}
	{\cal F}(q)={\cal F}(0)e^{-a\sqrt{q}}=\sigma_{bfpp}e^{-a\sqrt{q}}.
\end{eqnarray}

In the equation above, ${\cal F}(0)$ corresponds to the function's value at $q=0$, which equals the total BFPP cross section. Unlike in our previous calculations \cite{14}, we find here that the behavior of $\mathcal{F}(q)$ is best described by the functional form $e^{-a\sqrt{q}}$. The constant $a$ is independent of both the charge and the energy of the heavy ions. Consequently, the BFPP probability as a function of impact parameter can be expressed as

\begin{equation} \label{17}
	P_{bfpp}(b)=\frac{1}{2\pi b} \frac{d\sigma_{bfpp}}{d b}=\frac{1}{2\pi}\mathcal{F}(0)\int_0^{\infty} qdq \, J_0(qb) \, e^{-a\sqrt{q}}.
\end{equation}

For bound-free $e^{+}e^{-}$ pair production, the constant $a$ takes the values $3.468 \lambda_{C}^{e}$ and $3.599 \lambda_{C}^{e}$ for $Pb+Pb$ collisions at $\gamma=1500$ and $\gamma=3000$, respectively. It is proportional to the Compton wavelength of the electron ($\lambda_{C}^{e}$) and independent of both the collision energy and the nuclear charge. The quantity
$\mathcal{F}(0)$ corresponds to the total cross section at $q=0$. Using the Monte Carlo integration method, this yields $\sigma_{bfpp}=181 barn$ at $\gamma=1500$ and $199 barn$ at $\gamma=3000$. Finally, $P_{bfpp}(b)$ denotes the probability of producing a single bound-free $e^{+}e^{-}$ pair from second-order Feynman diagrams.

Once the free and bound-free pair production probabilities have been formulated, the corresponding cross sections can be obtained by integrating over the impact parameter in Eq.~(\ref{3}) for relativistic $Pb+Pb$ collisions at the LHC, with Lorentz factors of $\gamma=1500$ and $\gamma=3000$.
	
	\section{Results and Comments}
	
In ultra-relativistic $Pb+Pb$ collisions at Lorentz factors of $\gamma=1500$ and $\gamma=3000$, the calculated cross sections for free $\mu^{+}\mu^{-}$ pair production and bound-free $e^{+}e^{-}$ pair production are $1.64 mbarn$ and $2.23 mbarn$,respectively. These cross sections were computed within the framework of QED perturbation theory.  The functions $\mathcal{F}_1(q_1)$ and $\mathcal{F}(q)$ were evaluated using Monte Carlo integration, with the multi-dimensional integrals sampled at approximately 10 million randomly chosen points to ensure convergence to the theoretical values. The associated numerical uncertainty in these calculations is estimated to be  5 ~\%{} or less.
We define the single-pair production probabilities  $P_{\mu^{+}\mu^{-}}(b)$ and $P_{bfpp}(b)$ as functions of the impact parameter for the free $\mu^{+}\mu^{-}$ and bound-free $e^{+}e^{-}$ pair production processes. Our calculated cross sections for free $\mu^{+}\mu^{-}$ and bound-free $e^{+}e^{-}$ pair production in $Pb+Pb$ collisions have been compared with the results reported in \cite{15} and show good agreement. The values from \cite{15} that calculated by using Equivalent Photon Approximation (EPA) are $1.8 mbarn$ at $\gamma=1500$ and $2.1 mbarn$ at $\gamma=3000$. The discrepancy between our results and those obtained via the EPA in \cite{15} is approximately 8 ~\%{} and 6 ~\%{} for $\gamma=1500$ and $\gamma=3000$, respectively. For cross sections calculated using first-order perturbation theory, \cite{15} reports values of $2.2 mbarn$ at $\gamma=1500$ and $2.6 mbarn$ at $\gamma=3000$. Compared to these, our results differ by about 25 ~\%{} and 14 ~\%{}  for the respective Lorentz factors. Furthermore, when comparing our impact parameter dependent probability function with the approximated functions used in \cite{15} for calculating multiple pair production, our results align closely with both the first-order perturbative and the EPA approaches presented there. In particular, the functional behavior of the pair production probability versus impact parameter more closely resembles the EPA approximation, which employs a more sensitive calculation. Consequently, our final cross-section values are also in closer quantitative agreement with the EPA results.

In the future, we hope these cross sections are expected to provide new insights into QED in strong electromagnetic fields and to extend our understanding through further studies of multiple pair production.

\end{document}